\documentclass[%
reprint,pra,
amsmath,amssymb,
aps
]{revtex4-1}

\usepackage{graphicx}
\usepackage{dcolumn}
\usepackage{bm}

\begin{document}

\title{Non-Hermitian Majorana modes protect degenerate steady states}

\author{Simon Lieu}

\affiliation{
   Blackett Laboratory, Imperial College London, London SW7 2AZ, U.K.
   }
\affiliation{Cavendish Laboratory, University of Cambridge, JJ Thomson Avenue, Cambridge, CB3 0HE, U.K.}

\date{\today}
\begin{abstract}

We introduce non-Hermitian generalizations of Majorana zero modes (MZMs) which appear in the topological phase of a weakly dissipative Kitaev chain coupled to a Markovian bath. Notably, the presence of MZMs ensures  that the steady state in the absence of decoherence events is  two-fold degenerate.  Within a stochastic wavefunction approach, the effective Hamiltonian governing the coherent, non-unitary dynamics retains BDI classification of the closed limit, but belongs to one of four non-Hermitian ``flavors'' of the ten-fold way. We argue for the stability of MZMs due to a  generalization of particle-hole symmetry, and uncover the resulting topological phase diagram. Qualitative features of our study generalize to two-dimensional chiral superconductors. The dissipative superconducting chain can be mapped to an Ising model in  a complex transverse field, and  we discuss potential signatures of the degeneracy.
\end{abstract}

\maketitle

\section{Introduction}

Recent experimental \cite{zhou2018, weimann2017, xue2017,malzard2018,rudner2015, poli2015} and theoretical \cite{shen2018a, gong2018,kunst2018,yao2018a,yao2018b,ezawa2019,malzard2015, budich2018,klett2017,menke2017,yin2018, mcclarty2019,kawabata2018,thomale2019,jin2019,yuce2019,herviou2019,pikulin2013,borgnia2019,lieu2018a, lieu2018b} efforts have aimed at generalizing  topological band theory to complex band structures which dictate the dynamics of open systems.  While origins of this field primarily began with motivations  from classical topological photonics in the presence of gain or loss, more recent studies suggest that the formalism is very relevant for quantum many-body physics \cite{zyuzin2019,yoshida2018, kozii2017, shen2018b, lieu2018b, peano2016, mcdonald2018, wang2019,barnett2015, katsura2018, moos2019, choi2019}.

In this work, we use the framework of non-Hermitian topology to demonstrate that a dissipative quantum many-body system can possess symmetry-protected topologically-degenerate steady states, in analogy with the equilibrium paradigm of topologically-degenerate fermionic ground states. In the topological phase of superconductors, orthogonal ground states are related to each other by a nonlocal zero-energy excitation composed of a pair of Majoranas: $\left| \text{gnd}_2  \right\rangle = \beta_0^\dagger \left| \text{gnd}_1  \right\rangle $ where $\beta_0^\dagger=\alpha_l + i \alpha_r$ and $\alpha_{l/r}$ are Majorana fermions on the left/right boundary of the sample \cite{kitaev2001}. In open quantum systems, there is no notion of a ground state, but systems generically evolve into a steady state. We demonstrate that the steady state of an open system can be non-unique due to non-Hermitian Majorana zero modes which connect orthogonal states. 

The ten-fold way \cite{ryu2010} classifies symmetry-protected topological phases of free fermions in Hermitian systems. In a non-Hermitian setting there are more than ten  random matrix ensembles, called the  Bernard-LeClair (BL) classes \cite{bernard2002} which generalize the Altland-Zirnbauer (AZ) classes \cite{altland1997} in the absence of Hermiticity. Breaking Hermiticity leads to a larger number of symmetry classes since the Hamiltonian matrix obeys $H \neq H^\dagger \implies H^* \neq H^T$ such that time-reversal and particle-hole symmetry can be represented in two inequivalent ways, depending on whether the symmetry involves conjugation or transposition of the matrix.

Recent studies have constructed a topological periodic table using the BL classes as a basis, called the 38-fold way \cite{kawabata2018BL, zhou2018BL}. In this work, we discuss a physical system which hosts edge modes which are protected by BL symmetries.  Specifically, we study a non-Hermitian  superconductor with particle-hole symmetry whose Hamiltonian obeys: $H_k = -  \Sigma H_{-k}^T \Sigma$ but explicitly breaks the conventional expression: $H_k \neq - \Sigma H_{-k}^*\Sigma$ where $\Sigma$ transforms particles to holes.  We  show that this non-Hermitian particle-hole symmetry can be used to protect topological Majorana modes in both 1D and 2D, leading to robustly degenerate steady states.

\section{Four flavors of the ten-fold way}

Before specializing to the model, we uncover four different ``flavors'' of the ten-fold way which can arise in non-Hermitian models. In the Hermitian limit, the first-quantized Hamiltonian (matrix)  $\mathcal{H}=\sum_{i,j}H_{i,j}c_i^\dagger c_j$ can be classified according to the following symmetries
\begin{subequations}
\begin{align}
\text{TRS}_c:\qquad H & = U_T H^{*}U_T^{\dagger},\qquad U_T U_T^{*}=\pm\mathbb{I} \label{eq:trs}\\
\text{PHS}_c:\qquad H & = -U_C H^{*}U_C^{\dagger},\qquad U_C U_C^{*}=\pm\mathbb{I}\label{eq:phs}\\
\text{chiral}:\qquad H & = - U_S HU_S^{\dagger},\qquad U_S^2=\mathbb{I}. \label{eq:chiral}
\end{align}
\end{subequations}
This leads to $3 \times 3+1=10$ distinct classes, since there are three options for TRS (no TRS, $U_T U_T^{*}=\mathbb{I}$, or $U_T U_T^{*}=-\mathbb{I}$), three options for PHS,  and chiral symmetry can exist in the absence of both \cite{ryu2010}.

The crucial insight of Bernard and LeClair \cite{bernard2002} was to note that $H^T \neq H^*$ for non-Hermitian models (by definition). Thus once non-Hermitian terms are included, $H$ can still possess TRS and/or PHS but they can be represented in two distinct ways, i.e.~ via transposition ($t$) or conjugation ($c$). We note that this leads to four distinct flavors of the ten-fold way, enumerated below. If both TRS and PHS involve transposition, then
\begin{subequations}
\begin{align}
\text{TRS}_t:\qquad H & = U_T H^{T}U_T^{\dagger},\qquad U_T U_T^{*}=\pm\mathbb{I} \label{eq:trs}\\
\text{PHS}_t:\qquad H & = -U_C H^{T}U_C^{\dagger},\qquad U_C U_C^{*}=\pm\mathbb{I}\label{eq:phs}\\
\text{chiral}:\qquad H & = - U_S HU_S^{\dagger},\qquad U_S^2=\mathbb{I},\label{eq:chiral}
\end{align}
\end{subequations}
generate a distinct ten-fold way.  Mixing transposition and conjugation leads to
\begin{subequations}
\begin{align}
\text{TRS}_c:\qquad H & = U_T H^{*}U_T^{\dagger},\qquad U_T U_T^{*}=\pm\mathbb{I} \label{eq:trs}\\
\text{PHS}_t:\qquad H & = -U_C H^{T}U_C^{\dagger},\qquad U_C U_C^{*}=\pm\mathbb{I}\label{eq:phs}\\
\text{PAH}:\qquad H & = - U_S H^\dagger U_S^{\dagger},\qquad U_S^2=\mathbb{I},\label{eq:chiral}
\end{align}
\end{subequations}
and
\begin{subequations}
\begin{align}
\text{TRS}_t:\qquad H & = U_T H^{T}U_T^{\dagger},\qquad U_T U_T^{*}=\pm\mathbb{I} \label{eq:trs}\\
\text{PHS}_c:\qquad H & = -U_C H^{*}U_C^{\dagger},\qquad U_C U_C^{*}=\pm\mathbb{I}\label{eq:phs}\\
\text{PAH}:\qquad H & = - U_S H^\dagger U_S^{\dagger},\qquad U_S^2=\mathbb{I},\label{eq:chiral}
\end{align}
\end{subequations}
where pseudo-anti-Hermiticity (PAH) generalizes chiral symmetry and is guaranteed if the model has both TRS and PHS of mixed $t,c$ character. The last two sets are called classes ``AZ'' and ``AZ$^\dagger$'' in Ref.~\cite{kawabata2018BL}. (Note that there are many redundancies between the $4 \times 10=40$ classes listed above, e.g.~we have counted the symmetry-less class A four times.)

\begin{figure}
\begin{centering}
\includegraphics[scale=0.2]{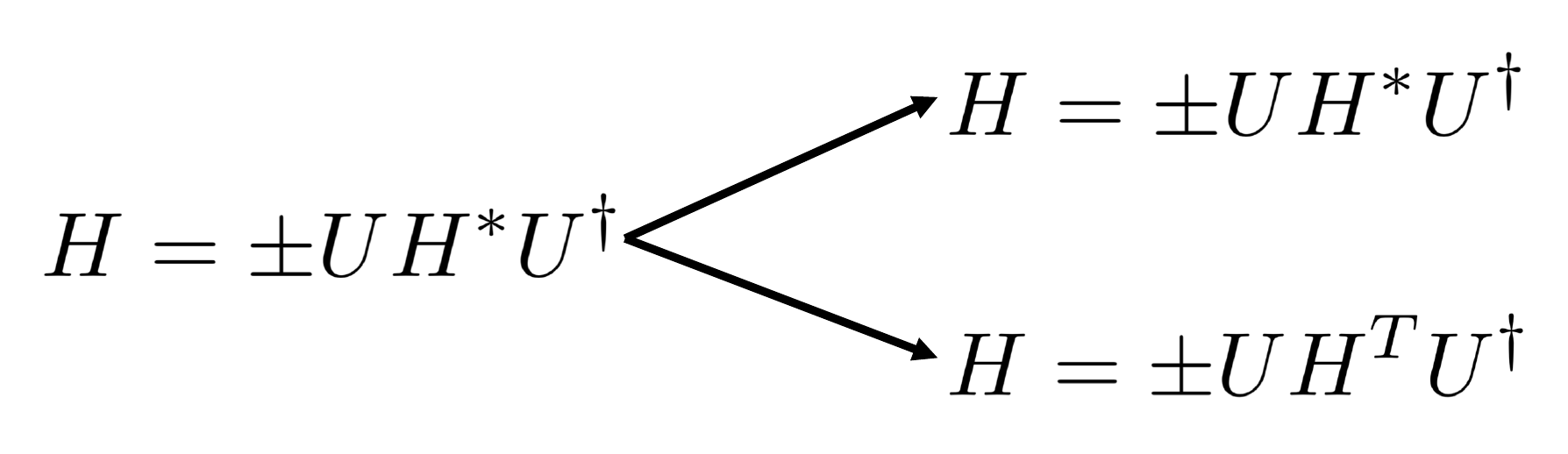}
\par\end{centering}
\caption{\label{fig:symSplit}  A Hermitian Hamiltonian with TRS and/or PHS (left) can preserve its symmetries once non-Hermitian terms are added, but the corresponding expression will either involve conjugation or transposition (right) which are inequivalent operations. This generates $2 \times 2=4$ flavors of the ten-fold way.}
\end{figure}

In the Hermitian limit $H=H^\dagger$, the four sets of symmetries defined above are identical and hence they are only distinct in the presence of non-Hermitian terms. Adding non-Hermitian terms to the model $H$ can preserve ten-fold way symmetries, but will generically move the system to one of the four sets given above. This is sketched in Fig.~\ref{fig:symSplit}. While these classes are a subset of the 38 topologically-distinct non-Hermitian symmetry classes as defined in Refs.~\cite{kawabata2018BL, zhou2018BL}, they are physically motivated as natural dissipative extensions of the fermionic ten-fold way. In this work, we discuss a physical scenario where the ten-fold way generated by $\text{TRS}_t$ and $\text{PHS}_t$ arises naturally and leads to non-Hermitian generalizations of gapless Majorana modes. 

\section{Setup, model, and symmetries}

We study the non-unitary time evolution  of a topological superconductor (TSC) coupled to a Markovian bath  which is allowed to remove electrons from the system, in the limit where removal events are rare. Our starting point is the Lindblad master equation
\begin{equation} \label{eq:lind}
\frac{d \rho}{dt}= \mathcal{L}(\rho)= -i\left( \mathcal{H}_{\text{eff}} \rho - \rho \mathcal{H}_{\text{eff}}^\dagger   \right) + 2 \sum_j \gamma_j L_j \rho L_j^\dagger
\end{equation}
where $\rho$ is the density matrix of the  system, $L_j$ are the Lindblad jump operators which represent decoherence events (occurring at rate $\gamma_j$), and  $\mathcal{H}_{\text{eff}} = \mathcal{H}- i \sum_j \gamma_j L_j^\dagger L_j$ such that $\mathcal{H}$ generates the unitary part of the evolution \cite{lindblad1976}. $\mathcal{L}$ is a non-Hermitian ``superoperator'' which generates the non-unitary dynamics.

The Lindblad master equation \eqref{eq:lind}  lends itself to a convenient physical interpretation known as the quantum stochastic wavefunction approach \cite{plenio1998, daley2014}: In a time step $dt$, a system prepared in a pure state will either evolve coherently according to the non-Hermitian effective Hamiltonian $\mathcal{H}_{\text{eff}}$, or a ``quantum jump event'' will decohere the system by moving a pure state from $| \psi \rangle$ to $L_j | \psi \rangle$. Averaging over all such trajectories will produce the same expectation values as formally solving the Lindblad master equation for the evolution of the density matrix.

We specialize to the case when the jump events occur at a low rate such that the system's dynamics can be reasonably modeled by studying the coherent, non-unitary evolution generated by $\mathcal{H}_{\text{eff}}$.  (This approximation is exact in the limit of no jump events; see Sec.~\ref{sec:tfim}.) Specifically, we begin by studying a fermionic Kitaev chain Hamiltonian coupled to a bath which can remove electrons at each lattice site: $L_n=c_n$, which leads to the effective Hamiltonian

\begin{equation} \label{eq:tscham}
\mathcal{H}_{c}= 2u \sum_{n=1}^N c_n^\dagger c_n - \Delta \sum_{n=1}^N \left( c_n^\dagger c_{n+1} + c_n^\dagger c_{n+1}^\dagger +h.c. \right)
\end{equation}
where $u=\mu -i \gamma \in \mathbb{C}$, $\Delta \in \mathbb{R}$. The terms represent an equilibrium chemical potential $\mu$, dissipation which occurs at a uniform rate $\gamma$ across lattice sites, and equal hopping and pairing strength $\Delta$. Mathematically, this is  a Kitaev chain with a complex potential. The many-body spectrum is now complex with negative imaginary components which represents ``probability leakage'' into the environment. The steady state of the system is defined as the eigenstate with the largest (least negative) imaginary energy.

\begin{figure}
\begin{centering}
\includegraphics[scale=0.3]{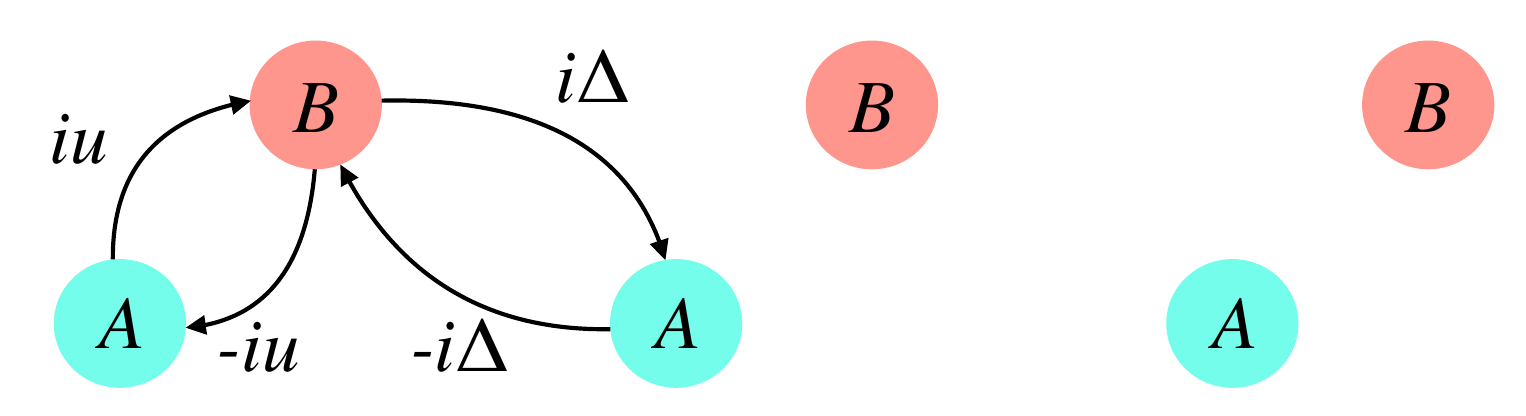}
\par\end{centering}
\caption{\label{fig:sshMaj} Dissipative Kitaev chain effective Hamiltonian in a Majorana basis \eqref{eq:majham}. The model transforms to a  Su-Schrieffer-Heeger model with non-Hermitian intracell hopping $\pm iu \in \mathbb{C}$, and Hermitian intercell hopping $\pm i \Delta \in \mathbb{I}$.}
\end{figure}

To uncover the symmetries of the Hamiltonian it is useful to  transform the model to a Majorana basis. Making the replacement: $\alpha_{i,A}=c_i+c_i^\dagger, \alpha_{i,B}=i(c_i-c_i^\dagger)$ leads to
\begin{align} \label{eq:majham}
\mathcal{H}_{\alpha} &= i u \sum_{n=1}^N \left( \alpha_{i,A} \alpha_{i,B} - \alpha_{i,B} \alpha_{i,A} \right) \\ &+ i \Delta  \sum_{n=1}^{N-1} \left( \alpha_{i,B} \alpha_{i+1,A} -  \alpha_{i+1,A}  \alpha_{i,B}  \right) \nonumber.
\end{align}
The Hamiltonian transforms to a Su-Schrieffer-Heeger (SSH) model \cite{su1979} with hopping phases which break Hermiticity, depicted in Fig.~\ref{fig:sshMaj}. (This is the chiral-SSH model, previously studied in Ref.~\citep{lieu2018a}.)

The Hermitian Kitaev chain belongs to class BDI within the ten-fold way, leading to a $\mathbb{Z}$ classification which suggests that an arbitrary number of edge modes can be protected due to the chiral symmetry in a Majorana basis. In the absence of Hermiticity, the model again falls into class BDI, but this time it belongs to the ten-fold way generated by $\text{TRS}_{t}$ and $\text{PHS}_{t}$. More specifically, the  effective Hamiltonian in a Majorana basis satisfies the following symmetries: $H_{\alpha} = -H_{\alpha}^T, H_{\alpha} = \tau_z H_{\alpha}^T \tau_z, H_{\alpha} = -\tau_z H_{\alpha} \tau_z$ where $\tau_z=\mathbb{I}_N \otimes \sigma_z$ and $\sigma_z$ is the Pauli matrix. The Hamiltonian keeps its TRS and PHS symmetries, but both are represented via transposition rather than conjugation.  Naturally, the Hamiltonian in a complex fermion basis acquires the same three symmetries, represented as: $H_c = - \Sigma_x H_c^T \Sigma_x, H_c = H_c^T, H_c= - \Sigma_x H_c  \Sigma_x$ where $\Sigma_x= \sigma_x \otimes \mathbb{I}_N$.

To diagonalize the Hamiltonian, we make a transformation to fermionic quasiparticles
\begin{equation}
\left(\begin{array}{cc} \mathbf{\beta} & \mathbf{\bar{\beta}}^\dagger \end{array}\right) = \left(\begin{array}{cc} \mathbf{c}^\dagger & \mathbf{c} \end{array}\right) V, \qquad 
\left(\begin{array}{c} \mathbf{\bar{\beta}}^\dagger \\ \mathbf{\beta} \end{array}\right) = V^{-1} \left(\begin{array}{c} \mathbf{c} \\ \mathbf{c}^\dagger \end{array}\right).
\end{equation}
The two types of quasiparticles $\beta,\bar{\beta}^\dagger$ arise due to the right and left eigenvectors of non-Hermitian matrices \cite{kawabata2018}. The particle-hole symmetry of the Hamiltonian imposes a structure on the transformation matrix
\begin{equation}
 V = \Sigma_x \left( V^{-1} \right)^T \Sigma_x.
\end{equation}
Remarkably, this expression guarantees that quasiparticles obey generalized fermionic statistics
\begin{equation}
\{ \beta_i, \bar{\beta}_j^{\dagger}  \}= \delta_{i,j}, \, \{ \bar{\beta}_i^{\dagger} , \bar{\beta}_j^{\dagger} \} = \{ \beta_i, \beta_j \} = 0.
\end{equation}
The diagonalized second-quantized Hamiltonian reads
\begin{equation}
\mathcal{H}_c =  \frac{1}{2}  
\left(\begin{array}{cc} \mathbf{\beta} & \mathbf{\bar{\beta}}^\dagger \end{array}\right) \Lambda \left(\begin{array}{c} \mathbf{\bar{\beta}}^\dagger \\ \mathbf{\beta} \end{array}\right) = \sum_i E_i \left( \bar{\beta}_i^\dagger \beta_i + \frac{1}{2} \right)
\end{equation}
where $\Lambda=\text{Diag}[-E_1,\ldots,-E_N, E_1,\ldots,E_N]$ is a diagonal matrix whose entries correspond to the energies of the system and  $\text{Re}[E_i]>0$. The quasiparticle vacuum state is defined as: $\beta_i  \left| \text{vac} \right\rangle =0$, and an excited state with energy $E_i$ is $\bar{\beta}_i^\dagger  \left| \text{vac} \right\rangle$.

\section{Stability of Majorana modes}
 
Before calculating the spectrum, we generalize the stability of Majorana zero modes to include robustness against non-Hermitian terms in the Hamiltonian. In the Hermitian Kitaev chain, the MZM is protected at zero energy due to its particle-hole symmetry
\begin{equation}
\psi_{\text{0}} \propto \Sigma_x \psi_{\text{0}}^* \implies E_{\text{0}}=0
\end{equation}
where $H_c \psi_{\text{0}}=0$. Any term entering the Hamiltonian which preserves the bandgap cannot perturb the MZM away from zero energy.

In the non-Hermitian case, each eigenvalue $E$ has an associated right and left eigenvector, defined as
\begin{align}
H_c \psi  &=E \psi   \\
H_c^{\dagger}\lambda  &=E^{*} \lambda .
\end{align}
The non-Hermitian MZM satisfies the condition
\begin{equation}
\psi_{\text{0}} \propto \Sigma_x \lambda_{\text{0}}^* \implies E_{\text{0}}=0.
\end{equation}
We find that the particle-hole symmetry: $H_c = - \Sigma_x H_c^T \Sigma_x$ protects the MZM at zero energy in direct analogy with the Hermitian case. This ensures a two-fold  degeneracy in the many-body spectrum of the dynamical system.  Our analysis  generalizes the protection of MZMs in a TSC with respect to Hermiticity-breaking terms in the Hamiltonian. (Chiral symmetry also ensures gapless edge modes, but the argument above can protect a MZM at the edge  even when both TRS and chiral symmetry are broken.)

In the analysis of Refs.~\cite{kawabata2018BL, zhou2018BL}, the model has a $\mathbb{Z}$ point-gap classification in 1D. If TRS is broken then this reduces to a $\mathbb{Z}_2$ classification.  Bulk band gaps have to close in both the real and imaginary plane simultaneously in order to remove a Majorana mode pinned to $E_0=0$, in agreement with our adiabatic argument.

\section{Many-body spectrum}

\begin{figure}
\begin{centering}
\includegraphics[scale=0.15]{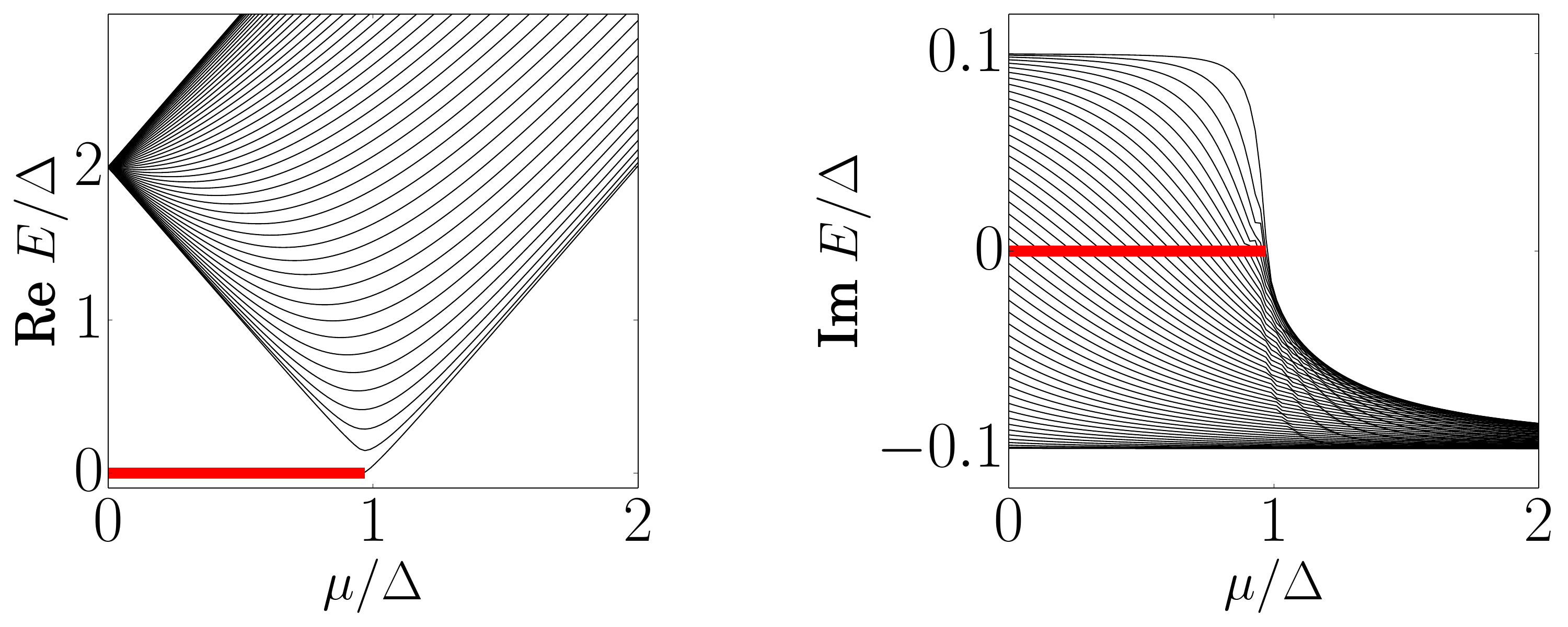}
\par\end{centering}
\caption{\label{fig:1dspec} Spectrum of the 1D non-Hermitian Kitaev chain, $\gamma/\Delta=0.1$. (Each eigenvalue comes in $\pm$ pairs; we only plot the positive-real-energy branch of the quasiparticle spectrum for clarity.) MZMs (red) exist on opposite ends of the spectrum and couple to form a quasiparticle zero-mode. Bulk modes (black) are generically complex. Positive imaginary modes represent amplifying hole bands, while negative modes represent decaying electronic bands.}
\end{figure}

The spectrum of the system for weak decay is given in Fig.~\ref{fig:1dspec}. We find that the energy of the MZM remains unchanged upon inclusion of decay, in agreement with the analysis from the previous section. This is not true for bulk modes, all of which acquire a non-zero imaginary component to their energy. Interestingly, we find that some quasiparticles get amplified while others decay.  We can understand this behavior by examining the composition of the quasiparticles in terms of electrons and holes. Quasiparticles which are mostly composed of electrons ($c^\dagger$ terms) acquire a negative imaginary energy indicating decay, while modes composed of holes ($c$ terms) acquire a positive imaginary energy indicating growth. Intuitively this agrees with the idea that electronic dissipation leads to the proliferation of holes.  MZMs are equally composed of electrons and holes, hence their imaginary component is  zero. (Note that in order to properly calculate observables we must renormalize the wavefunction after time evolving \cite{plenio1998, daley2014}.)

What does a complex energy spectrum imply for the dynamics of the system? An arbitrary initial state can be rewritten as a superposition of quasiparticles $\bar{\beta}_j^\dagger$ acting on the vacuum
\begin{equation}
\left| \Psi \right\rangle = \sum_{ \{ i \} } \kappa_{ \{ i \}} \Pi_{ j \in \{ i \} } \bar{\beta}_j^\dagger \left| \text{vac} \right\rangle
\end{equation}
where $\{i\}$ represents all $2^N$ permutations of occupied states, and $\kappa_{ \{ i \}}$ is the associated amplitude. Upon evolving the state in time by $\exp(-i \mathcal{H} t)$, the terms with the most hole-like quasiparticles will start to dominate the wavefunction, since these are the modes which decay least as a function of time. The steady state of the system is $\left| \text{SS}_1 \right\rangle = \Pi_{ \{i\}'} \bar{\beta}_i^\dagger \left| \text{vac} \right\rangle $ where the set $\{i\}'$ includes all quasiparticles with $\text{Im}[E_i]>0$. This  state is degenerate in the topological phase due to MZMs which relate the steady states to each other $\left| \text{SS}_2 \right\rangle =\bar{\beta}_0^\dagger \left| \text{SS}_1 \right\rangle$.

\begin{figure}
\begin{centering}
\includegraphics[scale=0.4]{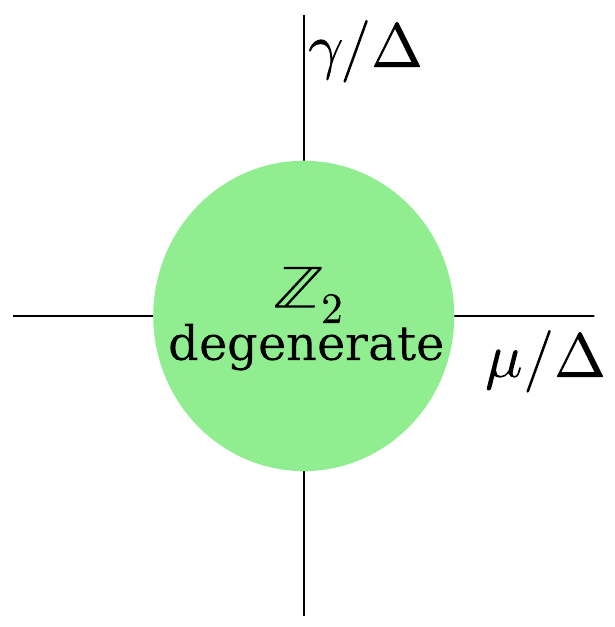}
\par\end{centering}
\caption{\label{fig:z2pd} Phase diagram of the dissipative Kitaev chain with chemical potential $\mu$, hopping and pairing strength $\Delta$, and dissipation  $\gamma$.  The green unit circle indicates a region where all eigenstates are two-fold degenerate due to the presence of a MZM on each edge.}
\end{figure}

\section{Dissipative phase diagram} We have seen that the presence of weak decay (small $\gamma$) leaves the Majorana mode pinned at zero energy while bulk modes generically pick up a complex dispersion. As the decay rate is further increased, it is possible to induce a topological phase transition by closing a band gap. In order to uncover the topological phase boundary, it is easiest to examine the Bloch Hamiltonian by transforming to momentum coordinates $k$. The bulk dispersion is found to be
\begin{equation}
E_k = \pm 2 \sqrt{ \left( \mu -i \gamma-\Delta \cos k \right)^2 +\Delta^2 \sin^2 k }.
\end{equation}
We easily identify the band closing points which occur at the critical values: $ \mu_c^2 + \gamma_c^2= \Delta_c^2$. The phase diagram is depicted in Fig.~\ref{fig:z2pd}. This agrees with our intuition: The Hamiltonian  maps to a non-Hermitian SSH model \eqref{eq:majham} which hosts edge modes only when the magnitude of the intracell hopping is smaller than that of the intercell hopping.

\section{Two-dimensional TSC}
\begin{figure}
\begin{centering}
\includegraphics[scale=0.15]{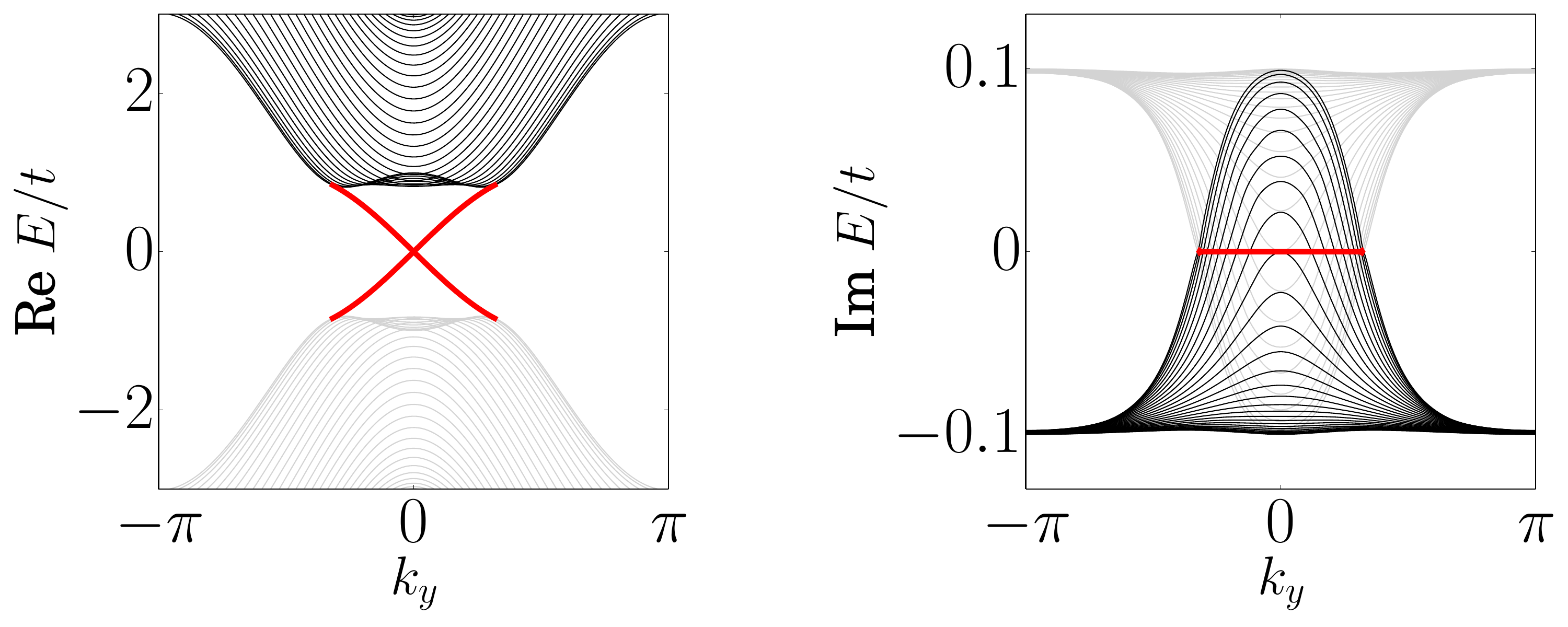}
\par\end{centering}
\caption{\label{fig:weaktsc} Semi-periodic spectrum in $k_y$ for the weakly-dissipative Hamiltonian given in Eq.~\eqref{eq:2dham} with parameters: $\Delta/t=0.5, \mu/t=1.0, \gamma/t=0.1$. Edge modes (red) retain their fully real dispersion, while bulk modes (black) are complex. The band crossing occurs at zero energy when $k=0$ due to MZMs which arise on opposite edges of the system. This is in agreement with symmetry considerations.}
\end{figure}

We demonstrate that the qualitative results from our study generalize to higher dimensions. Consider the class D superconductor in two-dimensions with a $\mathbb{Z}$ classification in the presence of uniform electronic loss $\gamma$
\begin{align}
\mathcal{H}_{\text{2D}} &= \sum_{m,n}  -t \left( c^\dagger_{m+1,n} c_{m,n} + c^\dagger_{m,n+1} c_{m,n} +h.c. \right) \notag \\ &+ \left( \Delta c_{m+1,n}^\dagger c_{m,n}^\dagger + h.c. \right) + \left( i \Delta c_{m,n+1}^\dagger c_{m,n}^\dagger + h.c. \right) \notag \\ &- \left( \mu +i \gamma -4t  \right) c_{m,n}^\dagger c_{m,n} \label{eq:2dham}
\end{align}
where we assume a square lattice geometry and $c_{m,n}$ annihilates a spinless fermion on lattice site $(m,n)$ of an $N \times N$ square lattice \cite{read2000}. The terms represent hopping $t$, pairing $\Delta$, and the chemical potential $\mu$. To observe edge modes we impose periodic boundary conditions in the $y$ direction while maintaining a finite slab in $x$. We  rewrite operators in terms of their Fourier transform
\begin{equation}
c_{m,n} = \frac{1}{\sqrt{N}} \sum_{k_y} e^{i k_y n} c_{m,k_y}.
\end{equation}
The Hamiltonian takes the form
\begin{equation}
\mathcal{H}_{\text{2D}} = \sum_{k_y \in (0,\pi)}  \mathbf{c}_{k_y}^\dagger H(k_y) \mathbf{c}_{k_y}
\end{equation}
where $\mathbf{c}_{k_y}=(c_{m=1,k_y}\ldots, c_{m=1,-k_y}^\dagger,\ldots)^T$. The spectrum is found by diagonalizing $H_{k_y}$ and is given in Fig.~\ref{fig:weaktsc} for a weakly decaying model. (Henceforth, we use the scalar $k$ to represent $k_y$.) Notice that  a single edge mode is localized on each side of the chain with opposite group velocity which results in a net chirality.

The MZM is protected at the high-symmetry point in the Brillouin zone due to particle-hole symmetry. The Bloch Hamiltonian satisfies
\begin{equation}
H(k) = -\Sigma_x H^T(-k) \Sigma_x.
\end{equation}
Non-Hermitian Majorana modes at a high-symmetry point in the Brillouin zone (e.g. $k=0$), are related via
\begin{equation}
 \psi_{\text{edge}}(k=0)  \propto \Sigma_x \lambda_{\text{edge}}^*(k=0).
\end{equation}
This again suggests that $E(k=0)=0$ due to particle-hole symmetry, in agreement with the numerics given in Fig.~\ref{fig:weaktsc}.

\section{Mapping to an Ising model in a complex field} \label{sec:tfim}

The coherent, non-unitary time evolution described in this work is exact in the absence of a ``quantum jump event,'' i.e. a process where an electron leaves the superconductor and enters the bath. In principle, one could continuously monitor the system to check if such an event has taken place. While this is difficult to do in electronic models, it has been achieved in spin systems coupled to a photonic cavity field, e.g.~spontaneous decay in an atomic level is accompanied by photonic emission which can be monitored. We describe how our analysis of the Hamiltonian in Eq.~\eqref{eq:tscham} predicts robustly degenerate  steady states  in a dissipative extension of the canonical transverse field Ising model (TFIM).

\begin{figure}
\begin{centering}
\includegraphics[scale=0.2]{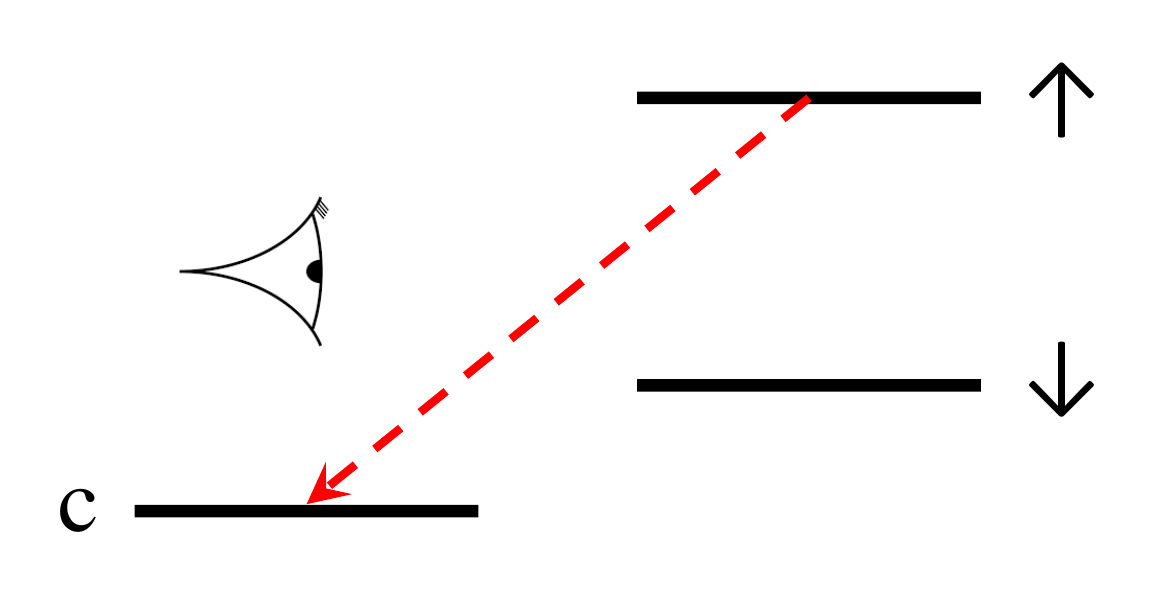}
\par\end{centering}
\caption{\label{fig:threeL} A two-level system $\uparrow, \downarrow$  can decay to  a third level $c$  by starting in $\uparrow$ and spontaneously emitting a photon into a cavity. (The $c$ state is a ``dark mode'' of the Lindbladian.) By checking for the absence of photonic emission, the three-level system is projected onto a two-level subspace which undergoes coherent, non-unitary dynamics. (Emission from $\uparrow$ to $\downarrow$ can be disallowed by selection rules.) }
\end{figure}

Before specializing to the many-body case, we first discuss a single atomic three-level system, with states: $\left| \uparrow \right\rangle , \left| \downarrow \right\rangle , \left| c \right\rangle $. We consider a scenario similar to Ref.~\cite{lee2014}: The two-level system $\left| \uparrow \right\rangle , \left| \downarrow \right\rangle$ undergoes unitary time evolution according to some Hermitian Hamiltonian $\mathcal{H}$. The only way for an atom to end up in the state $\left| c \right\rangle$ is by starting in the state $\left| \uparrow \right\rangle$ and spontaneously emitting a photon into a cavity mode. (See Fig.~\ref{fig:threeL}.) In the Lindblad formalism, the quantum jump operator of this dissipative three-level system is: $L = \left| c \right\rangle \left\langle \uparrow \right|$.

We suppose that we can constantly check if a photon has been emitted. (In practice this can be achieved via postselection by measuring the occupation of the $c$ state \cite{lee2014}.) Partially-projected spin systems have been experimentally probed for a single atom \cite{katz2006, sherman2013}, and many-body systems have been investigated theoretically \cite{lee2014, kozlowski2016}. Lack of emission leads to coherent, non-unitary dynamics generated by $\mathcal{H}_{\text{eff}}$ which represents exponential decay in the probability of finding the system in the $\left| \uparrow \right\rangle$ state as a function of time. This agrees with our intuition: Conditioning on the fact that no photons are emitted, the probability of finding the spin system in the up state decreases over time.

Consider $N$ of such three-level systems, which interact coherently according to a transverse-field Ising model. Then in the absence of photonic emission, the $\uparrow, \downarrow$ subsystem evolves via the effective Hamiltonian
\begin{equation} \label{eq:tfimham}
\mathcal{H}=-\Delta \sum_n \sigma_n^x \sigma_{n+1}^x + \mu \sum_n \sigma_n^z-i \gamma \sum_n\left( \sigma_n^z +\mathbb{I} \right) 
\end{equation}
where $\sigma_n^i$ represents the  $i$th Pauli spin operator on the lattice site $n$ of $N$.  The first two terms represent the standard Hermitian TFIM, while the last term represents the non-Hermitian contribution responsible for non-unitary dynamics of the eigenstates. 

The many-body spectrum can be found exactly by performing the Jordan-Wigner transformation \cite{Coleman1} to fermionic degrees of freedom
\begin{equation}
\sigma_j^+ =\exp\left(-i \pi \sum_{k=1}^{j-1} c_k^\dagger c_k \right) c_j^\dagger,  \qquad\sigma_j^z = 2 c_j^\dagger c_j -1.
\end{equation}
The Hamiltonian (neglecting constants) reads
\begin{equation} \label{eq:tscham2}
\mathcal{H}= 2 u \sum_n c_n^\dagger c_n - \Delta \sum_{n} \left( c_n^\dagger c_{n+1} + c_n^\dagger c_{n+1}^\dagger +h.c. \right)
\end{equation}
where $u=\mu-i\gamma$, and $c_n$ represents a complex spinless fermion on site $n$.  This implies that  the Ising model in a complex field shares the same spectrum as the dissipative Kitaev chain Eq.~\eqref{eq:tscham} studied in this work. The $\mathbb{Z}_2$ topological degeneracy of the Kitaev chain translates to a  $\mathbb{Z}_2$ degeneracy due to symmetry-breaking in the Ising model.

We briefly outline an experimental setup which could diagnose signatures of the degeneracy. Consider a chain of spins which is initially prepared in a symmetry-broken state in the $x$ direction, i.e. $\left| \rightarrow \rightarrow \rightarrow \ldots \right>$. We then quench the system by allowing the spin-up state in $z$ to decay to a third level by emitting a photon into a cavity mode with rate $\gamma$, and potentially applying a transverse field $\mu$. After propagating the system in time, but before the first photonic emission into the cavity, we turn off the evolving Hamiltonian and make  measurements in the resulting steady state. We expect this steady state to remain symmetry-broken when  $\gamma^2 + \mu^2 < \Delta^2$, i.e. $\left< \sigma_x \right> \neq 0$, whilst a unique steady-state is chosen in the opposite limit, i.e. $\left< \sigma_x \right> = 0$. Our simple setup predicts a quantum phase transition  for a non-Hermitian extension of the famous transverse-field Ising model.

\section{Conclusions} 

We have found non-Hermitian generalizations of Majorana zero modes which appear in the effective Hamiltonian description of topological superconductors coupled to a Markovian bath.  We began by uncovering four non-Hermitian flavors of the ten-fold way: Time-reversal  and particle-hole symmetry of the Altland-Zirnbauer classification can both be represented either via transposition or conjugation, leading to $2\times2=4$ sets of symmetry classes. We then demonstrated that the set generated by TRS  and PHS transposition is relevant for the dissipative Kitaev chain.  In the presence of electronic loss to an environment, the effective Hamiltonian retains its BDI classification with particle-hole symmetry protecting gapless Majorana modes at the edge. These modes ensure that the pure steady state of $\mathcal{H}_{\text{eff}}$ is two-fold degenerate: $\left| \text{SS}_2 \right\rangle =\bar{\beta}_0^\dagger \left| \text{SS}_1 \right\rangle$, where $\bar{\beta}_0^\dagger$ is a quasiparticle composed of a pair of  Majorana modes, in  analogy with the ground state degeneracy of the closed model. Our work extends  ideas from topological band theory to non-equilibrium many-body setups which evolve non-unitarily in time.

We uncovered the phase diagram of the non-Hermitian Kitaev chain in terms of dissipation strength, and demonstrated how generalizations of MZMs can arise in two-dimensional class D superconductors. While the coherent dynamics generated by $\mathcal{H}_{\text{eff}}$ is a good approximation in the weakly-dissipative limit, we have discussed a partially-projected spin system which will evolve exactly according to this Hamiltonian: In the absence of spontaneous decay (heralded by a lack of photonic emission), the steady state of a dissipative transverse-field Ising model will exhibit the same two-fold degeneracy discussed in this work  due to the Jordan-Wigner mapping which ensures that the models share the same spectrum. 


\textit{Acknowledgments.}  I thank R.~Barnett, D.~K.~K.~Lee, M.~McGinley, and N.~R.~Cooper for useful discussions and comments. I acknowledge financial support from the  Imperial College President's Scholarship.

\textit{Note added.} After submitting this work for publication, a preprint appeared which  discusses non-Hermitian Majorana modes in a different physical context \cite{okuma2019}.

\bibliography{csymBib}
\bibliographystyle{apsrev4-1}

\end{document}